\documentclass[12pt]{iopart}

\usepackage{iopams}  

\usepackage{graphicx}
\usepackage{bm}

\begin{document}


\title{Magnetic resonance of rubidium atoms passing through a multi-layered transmission magnetic grating}

\author{Y Nagata\footnote[1]{Present address: Department of Physics, Tokyo University of Science,
Shinjuku, Tokyo 162-8601, Japan. Present email address: yugo.nagata@rs.tus.ac.jp}, S Kurokawa and A Hatakeyama}
\address{Department of Applied Physics, Tokyo University of Agriculture and Technology, Koganei, Tokyo 184-8588, Japan}
\ead{ynagata@cc.tuat.ac.jp and hatakeya@cc.tuat.ac.jp}

\vspace{10pt}
\begin{indented}
\item[]March 2017
\end{indented}

\begin{abstract}
We measured the magnetic resonance of rubidium atoms passing through periodic magnetic fields
generated by two types of multi-layered transmission magnetic grating.
One of the gratings reported here was assembled by stacking four layers of magnetic films
so that the direction of magnetization alternated at each level.
The other grating was assembled so that the magnetization at each level was aligned.
For both types of grating, the experimental results were in good agreement with our calculations.
We studied the feasibility of extending the frequency band of the grating 
and narrowing its resonance linewidth by performing calculations.
For magnetic resonance precision spectroscopy, we conclude that the multi-layered transmission magnetic grating
can generate periodic fields with narrower linewidths at higher frequencies when a larger number of layers is assembled at a shorter period length. Moreover,
the frequency band of this type of grating can potentially achieve frequencies of up to hundreds of PHz.
\end{abstract}

\vspace{2pc}

\submitto{\jpb}


\section{\label{sec:level1}Introduction}

Atomic resonance is usually induced by an electromagnetic wave, such as that generated by a laser or a microwave,
and is utilized for the spectroscopic study of atomic properties.
Such spectroscopic techniques have contributed toward the broad scientific field seen in the present day \cite{dem}.
This type of spectroscopy is of great importance and has a high scientific impact.

Atomic resonance, induced not by an electromagnetic wave but instead by a periodic static potential, was predicted and observed decades ago \cite{ok,hade,sd}.
The resonance of energetic highly charged ions passing through a crystalline periodic electric field
has been investigated at the energy levels of the X-ray region \cite{rce1, rce2}.
Recently, this technique was applied to the precision spectroscopy of Li-like uranium ions $\mathrm{U^{89+}}$
irradiated on a Si crystal to study quantum electrodynamic effects at the 2s energy level \cite{rce3}.

This type of technique can also be applied to the magnetic resonance of atoms using periodic magnetic structures, 
which were extensively studied for the manipulation of cold atoms \cite{add1,add2,add3,add4}.
Such resonance was first observed for the Zeeman sub-levels of rubidium (Rb) atoms in an Rb vapor cell,
on which a periodic magnetic field was applied using currents \cite{h1}.
The frequency experienced by the atoms is given by $f=v/a$, with $a$ denoting the period length of the field and $v$ the atomic velocity.
In this experiment, a resonance of around $100$~kHz was observed for $a=1$~mm and $v\sim 100$~m/s.
This resonance is the so-called ``motion-induced resonance" (MIR).
The relative linewidth of the resonance frequency ($\delta f/f$) depends on the number of periods of the magnetic field felt by the atoms $N$,
and its full width at half maximum (FWHM) is approximately $1/N$.

The magnetic resonance of atoms has been extensively investigated in the microwave region,
for example nuclear magnetic moments and hyperfine structures \cite{Ramsey, towns}.
MIR has advantages compared with the microwave technique.
The frequency is continuously and widely tunable by changing $v$.
In contrast, the microwave frequency strongly depends on the dimensions of the cavity and cannot be easily changed.
The field strength of a periodic magnetic field can be changed by up to a few tenths of T by selecting a ferromagnetic material
of residual magnetization $\mu_0 M \sim 1$~T, where $\mu_0$ and $M$ represent the vacuum permeability
and magnetization, respectively.
This strength is comparable to, or larger than, the magnetic field generated by a superconducting microwave frequency cavity,
which is limited by a type II magnetic quench field of, for example, 0.24~T for niobium (Nb) \cite{sc}.

In recent years, the magnetic transition of a $\mathrm{{}^{229m}Th}$ nucleus has been proposed as a candidate for a nuclear clock, 
with a relative linewidth as low as $\Delta E/E \sim 10^{-20}$ for a transition energy constrained to between 6.3 and 18.3~eV,
and an intensive search for these transitions is underway \cite{Wense}.
MIR, for example, can achieve a transition energy of 10~eV (2.4~PHz) for $v=2.4 \times 10^7$~m/s and $a=10$~nm.
If the frequency band of MIR is extended by shortening $a$ and the linewidth is narrowed considerably by a larger value of $N$,
MIR can be a strong spectroscopic technique for the measurement of magnetic resonance.

\begin{figure}[b]
\includegraphics[width=1.\linewidth]{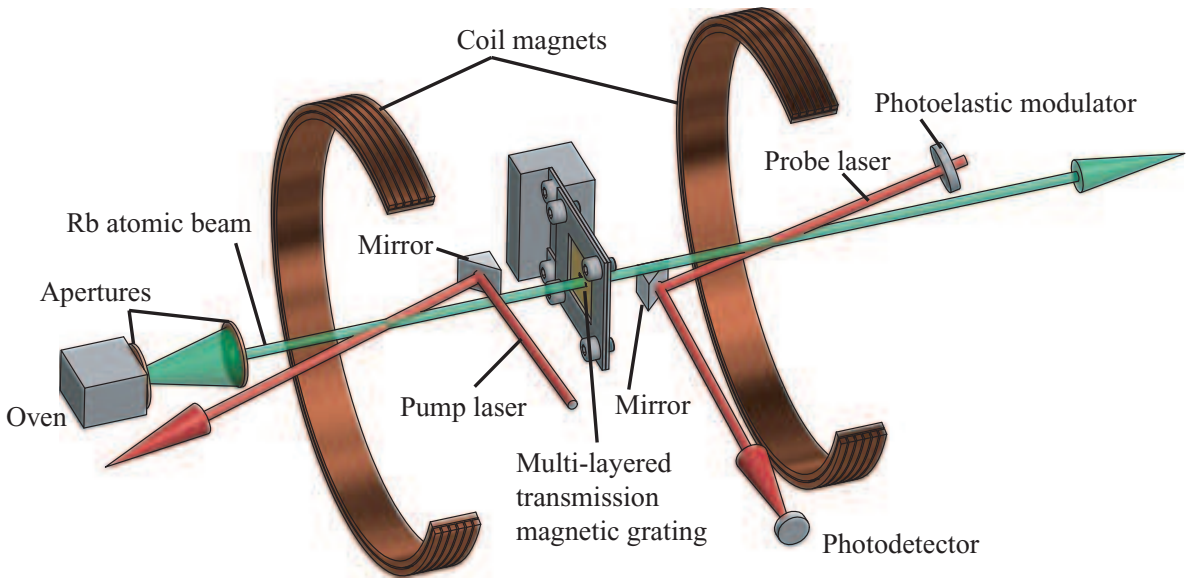}
\caption{\label{fig:setup}
Conceptual drawing of the experimental setup.
The Rb atomic beam is produced from the oven by collimation with two apertures
and passes through a multi-layered transmission magnetic grating.
The Rb atoms are polarized and detected by pump and probe lasers, respectively.
Two coil magnets generate a longitudinal magnetic field to lift the degeneracies of the Zeeman sub-levels of the Rb atoms.
}
\end{figure}

As an attempt to increase $N$, MIR was tested with an Rb atomic beam travelling between wire currents that
generated a periodic magnetic field for $N=28$ \cite{h3} and 10 \cite{h4}. The relative linewidths 
were observed as 4\% and 10\%, respectively, as expected.
However, it is not easy to assemble wires with a period of much less than 1~mm to obtain a shorter $a$.
MIR for shorter $a$ was demonstrated by using a monolayer transmission magnetic grating with a thin magnetic film
having a 150~$\mu$m line-and-space pattern, to obtain higher frequency resonance \cite{h5}.
In this experiment, Rb atoms were injected into the grating at an incline from the grating surface
with an angle of 10 degrees
such that they experienced a few oscillating cycles during passage through the grating.
Although it is possible to significantly enhance the resonant frequency,
it is difficult to obtain a larger value for $N$.

To address these limitations, in this paper we demonstrate MIR of Rb atoms 
passing through a multi-layered transmission magnetic grating
that was assembled by stacking four layers of magnetic film on top of each other.
We compare the experimental results of MIR with the results of our calculations.
We then discuss
the feasibility of extending the frequency band and narrowing the linewidth with the grating.
Our study shows the grating
can generate periodic fields with narrower linewidths at higher frequencies by assembling a larger $N$ at a shorter $a$,
and can be applied to the MIR in terms of the spectroscopy of magnetic resonance with the frequency potentially achieving up to hundreds of PHz.

\begin{figure}[h]
\includegraphics[width=1.\linewidth]{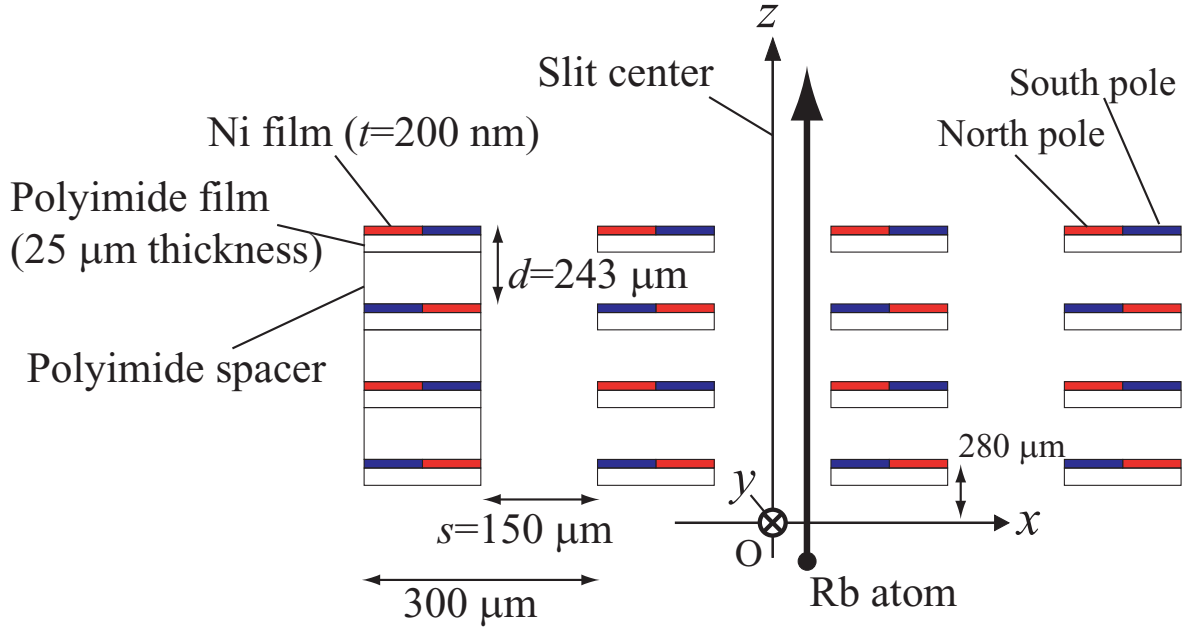}
\caption{\label{fig:grating}
The cross section of  part of a multi-layered transmission magnetic grating consisting of Ni films, polyimide films with 20 slits, and polyimide spacers.
The magnetization of the films alternates at each level.
}
\end{figure}

\section{\label{sec:level2}Experiment}

Figure~{\ref{fig:setup}} shows a conceptual drawing of our experimental setup,
which is the same as that used in a previous experiment \cite{h5} except for the replacement of a center grating.
Rb atoms were prepared in an oven heated up to a temperature of 473~K.
An Rb atomic beam was then produced by collimating Rb atoms emerging from the oven with two apertures, 
whose diameters and separation were 2~mm and 100~mm, respectively.
The flux of Rb atoms was expected to be $10^{18}$~${\mathrm{atoms/(s \cdot m^2)}}$.
Two coil magnets that generated a magnetic field $B_z$ up to 2~mT in the beam direction
were used to lift the degeneracies of the Zeeman sub-levels of the Rb atoms.
The Earth's magnetic field was cancelled in advance by the use of three axes of magnetic coils (not shown in the figure).
The Rb beam was polarized via
optical pumping by a circularly polarized pump laser with a wavelength of 780.24~nm through
the $F=3 \rightarrow F'=4$ transition of the $D_2$ line of ${}^{85}$Rb.
The power of the pump laser was about 100~${\mathrm{\mu W}}$, which was strong enough to obtain 
polarization signals with good signal-to-noise ratio.
The velocity of the Rb atoms was optically selected using the Doppler effect from the broad
velocity distribution of the thermal beam by laser frequency detuning with an acousto-optic modulator.
The Rb beam then passed through a multi-layered transmission magnetic grating.
The probe laser, which was also tuned to the $F=3 \rightarrow F'=4$ transition, was modulated between left- and right-hand circular polarizations
at 50~kHz by a photoelastic modulator, before being crossed with an Rb beam and detected by a photodetector.
The power of the probe laser was weaker than that of the pump laser and did not affect the signals.
If the frequency between Zeeman sub-levels $\gamma B_z$ coincides with $v/a$, MIR is induced and changes the absorption signal of the probe laser light,
where $\gamma$ is the gyromagnetic ratio and is $4.67$~GHz/T.
The modulation signal from the photodetector through a lock-in amplifier is proportional to the polarization of the $F=3$ Zeeman sub-levels in the direction
of the probe beam; this polarization is hereafter denoted as $\left<F_z \right>$.
By measuring the signal as a function of $B_z$ from 0~mT to 2~mT, MIR can then be observed.

The magnetic film was produced via vapor deposition of a nickel (Ni) layer with a thickness $t=200$~nm
on a 25-$\mathrm{\mu}$m-thick polyimide film
with 20 slits fabricated by means of chemical etching and with a slit size of width $s=150$~$\mathrm{\mu}$m and height $1$~mm
($y$ direction) at 300~$\mathrm{\mu}$m intervals, as shown in Fig.~{\ref{fig:grating}}.
Each magnetic film was magnetized parallel to the Ni film plane by applying a uniform magnetic field of 0.23~T,
which was strong enough to saturate the magnetization.
The residual magnetization $M$ was calculated from the B-H curve measured by a vibrating sample magnetometer
and $\mu_0 M$ = 280~mT.
The multi-layered transmission magnetic grating in Fig.~{\ref{fig:grating}} was assembled by stacking the magnetic films in four layers
so that the direction of magnetization alternated with each film, with polyimide sheet spacers inserted between them.
The average distance between layers $d$ was 243~$\mathrm{\mu}$m, and therefore $a=2d$.
Another multi-layered transmission magnetic grating was assembled so that the direction of magnetization of each of the films was aligned.
Such aligned magnetization has the advantage that the production procedure is considerably simplified, i.e.,
magnetic films can be magnetized at the same time after all magnetic films are assembled without the need for alternate stacking.
In this case, $a=d$.

\section{\label{sec:level3}Results and Discussion}

\begin{figure}[htb]
\includegraphics[width=1.\linewidth]{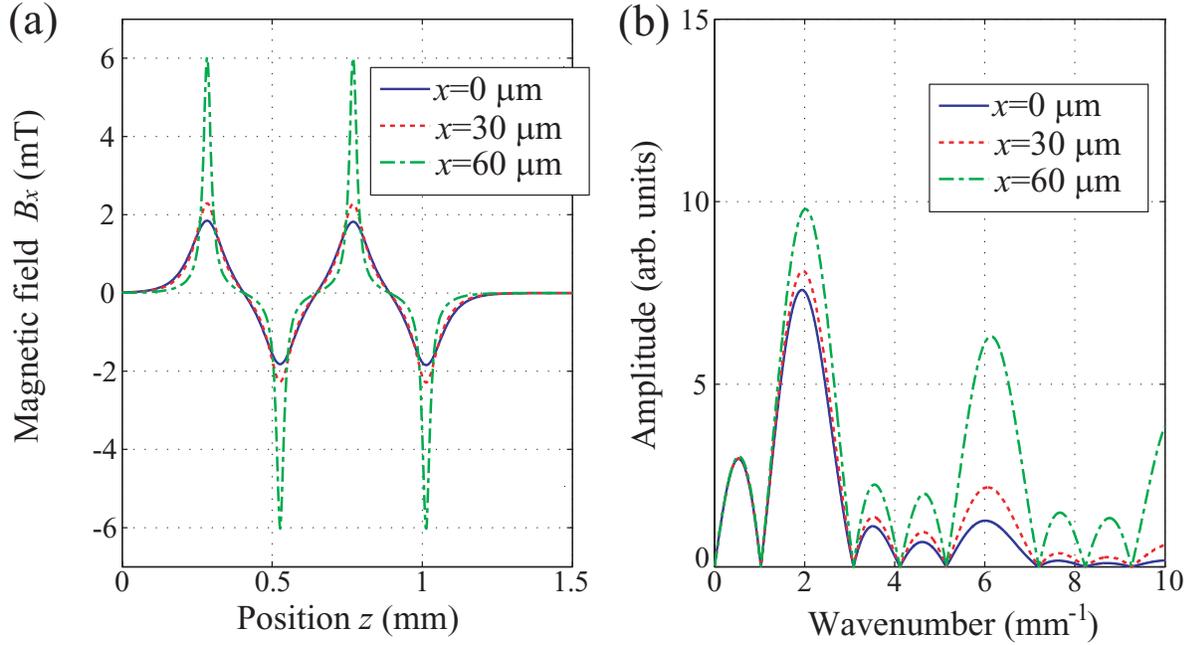}
\caption{\label{fig:2315G_50}
(a) The magnetic field felt by the Rb atoms passing through a multi-layered transmission magnetic grating with alternate magnetization.
Different lines show the $x$ dependence (see Fig.~{\ref{fig:grating}}).
(b) Fourier spectra calculated from the magnetic fields in (a).
}
\end{figure}

We calculated the $x$ component of the magnetic field $B_x$ along the path of the Rb atoms in a multi-layered transmission magnetic grating
for $\mu_0 M=280$~mT in alternate magnetization, as shown in Fig.~{\ref{fig:2315G_50}}~(a),
where we assumed that the height of the slit (the dimension in the $y$ direction) was infinity
and that the Rb atoms passed along the $z$ axis.
Blue solid, red dotted, and green dashed-dotted lines show the position $z$ dependences of magnetic fields
with the position from the slit center $x=0$, 30, and 60~$\mu$m, respectively (see Fig.~{\ref{fig:grating}}).
The number of periods $N$ was confirmed as 2.
As the paths of the Rb atoms lie close to the Ni film, the maximum magnetic field $B_{xm}$ becomes even stronger.
The $B_{xm}$ for $x = 60$~$\mu$m is three times larger than that for $x = 0$~$\mu$m.

Figure~{\ref{fig:2315G_50}}~(b) shows Fourier spectra calculated from the magnetic fields shown in Fig.~{\ref{fig:2315G_50}}~(a).
Different lines show the $x$ dependence.
The largest component appears at the frequency that corresponds to the first order Fourier component of $1/a = 2.1$~$\mathrm{mm^{-1}}$.
Although the magnetic fields in Fig.~{\ref{fig:2315G_50}}~(a) have different shapes depending on $x$,
each of the largest components appears at almost the same frequency.

\begin{figure}[h]
\includegraphics[width=1.\linewidth]{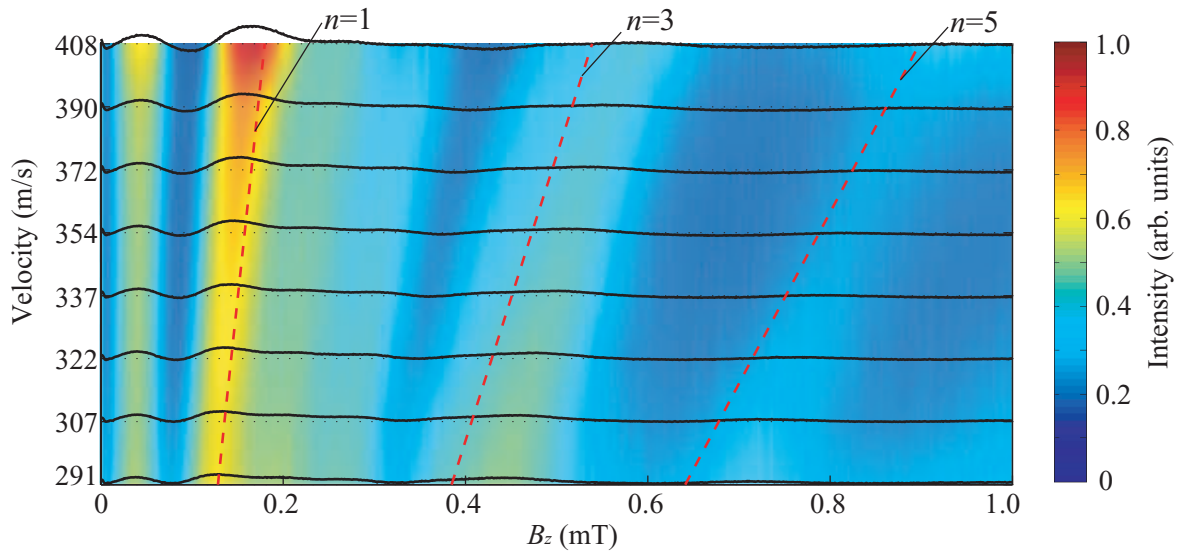}
\caption{\label{fig:detune}
The velocity dependence of MIR signals as a function of $B_z$.
}
\end{figure}

We measured the velocity dependence of MIR signals induced by a multi-layered transmission magnetic grating
with alternate magnetization as a function of $B_z$, as shown in Fig.~{\ref{fig:detune}}.
Black lines show experimental results from $v=291$ to 408~m/s
and are normalized by the number of atoms that depends on $v$ in the Maxwellian distribution of the atomic beam.
A color map shows the intensity of MIR signals interpolated between the black lines.
As $B_z$ is increased, the resonance frequency of the Rb atoms, $\gamma B_z$, becomes higher and thus coincides with $v/a$ for increased $v$.
To see this dependence, we draw a dotted line, indicated by the numbers ($n$) 1, 3, and 5, showing the resonance condition of $v=n \gamma B_z a$. 
The peaks of the experimental results generally follow the dotted lines well.

\begin{figure}[htb]
\includegraphics[width=0.8\linewidth]{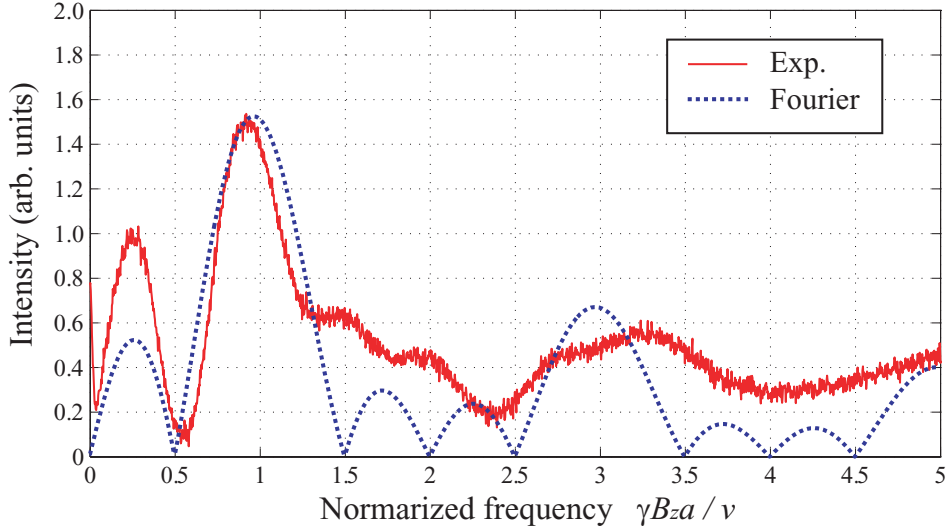}
\caption{\label{fig:results}
The MIR spectrum for $v=408$~m/s.
The red solid line shows the experimental result.
The horizontal axis shows the normalized frequency in units of $\gamma B_z a / v$.
The blue dotted line shows the calculated Fourier spectrum. 
}
\end{figure}

Figure~{\ref{fig:results}} shows a typical MIR spectrum selected from Fig.~{\ref{fig:detune}} for $v=408$~m/s.
The red solid line shows the experimental result.
The horizontal axis shows a normalized frequency $\gamma B_z a /v$ so that the frequency of the first order Fourier component
corresponds to 1.
We obtained the largest peak at around 1 and the FWHM was around 0.5, as expected.
The blue dotted line in Fig.~{\ref{fig:results}} shows a Fourier spectrum
averaged over the different $x$ from $-75$~$\mu$m to $75$~$\mu$m with 20 segments.
In the calculations, we assumed the atomic density was uniform across the slit and edge effects such as van der Waals forces were neglected.
The peak height of the first order Fourier component is normalized to the experimental result.
The calculation result is in acceptable agreement with the experimental result.

\begin{figure}[htb]
\includegraphics[width=1.\linewidth]{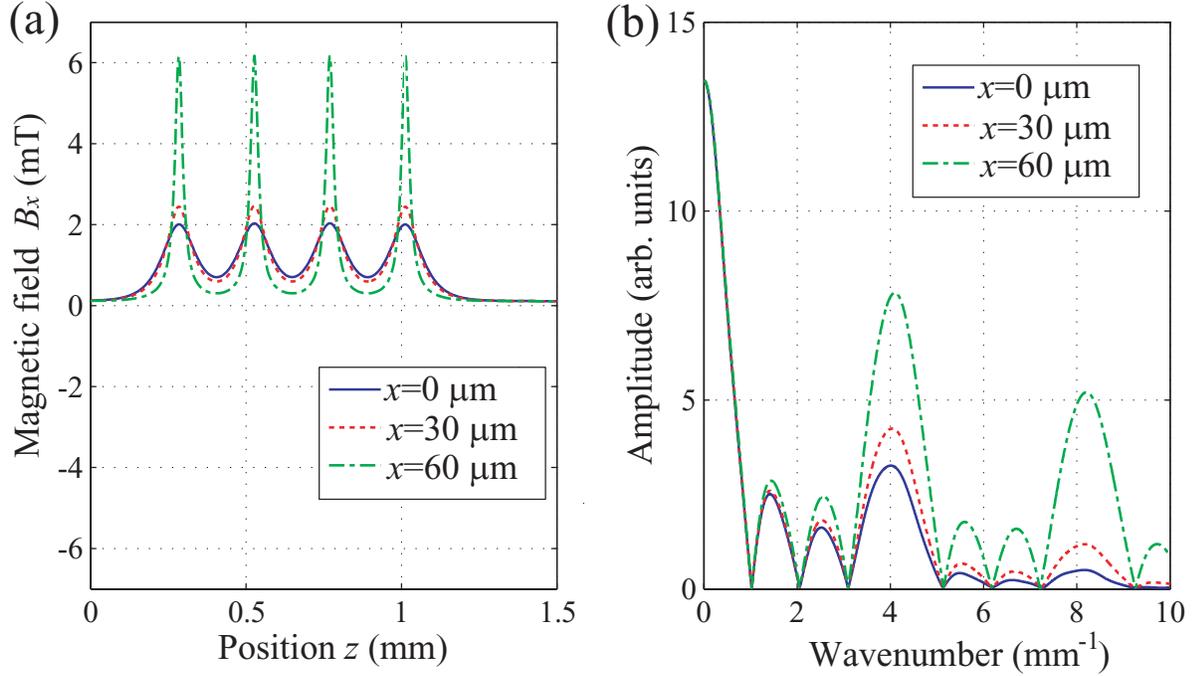}
\caption{\label{fig:2315G_502}
(a) The magnetic field felt by Rb atoms passing through a multi-layered transmission magnetic grating with aligned magnetization.
Different lines show the $x$ dependence.
(b) Fourier spectra calculated from the magnetic fields in (a).
}
\end{figure}

In the above experiment, four films were magnetized independently and stacked alternately.
If these films are stacked first and then magnetized at the same time,
the procedure for assembling the multi-layered transmission magnetic grating is simplified considerably.
Figures~{\ref{fig:2315G_502}}~(a) and (b) show the magnetic fields of this grating and the Fourier spectra, respectively,
while different lines show the $x$ dependence.
In contrast to Fig.~{\ref{fig:2315G_50}}~(a), the period length reduces by half ($a=d$).
As the magnetic fields are always positive,
the oscillation center of the magnetic field $B_b$ is biased by, for example, $B_b \sim 0.15$~mT at $x=0$~$\mathrm{\mu m}$.
$B_{xm}$ can be defined from $B_b$ and is roughly half, in comparison with the alternate magnetization.

The red solid line in Fig.~{\ref{fig:results2}} shows a MIR spectrum for $v=408$~m/s
with the multi-layered transmission magnetic grating in aligned magnetization.
The axes are the same as in Fig.~{\ref{fig:results}.
The blue dotted line shows a Fourier spectrum averaged over $x$ in the same way.
The Fourier calculation does not agree with the experimental result.
The largest peak in the experimental result appears at a frequency much smaller than 1.
Considering $B_b$, Zeeman splitting is changed from $\gamma B_z$ to $\gamma \sqrt{B_z^2 + B_b^2}$
or $\gamma B_z + \gamma \frac{B_b^2}{2B_z}$ for $B_b \ll B_z$.
For the first order resonance at $\sqrt{B_z^2 + B_b^2} =\frac{v}{\gamma a}=0.36$~mT,
$B_z \sim 0.33$~mT for $B_b \sim 0.15$~mT at $x=0$~$\mathrm{\mu m}$,
and is around 10\%
smaller than $0.36$~mT. 
Therefore this peak is shifted to the lower side.
To take into account $B_b$ for various $x$
and reproduce the detailed structure of the MIR spectrum, we solved a classical spin dynamics differential equation, 
\begin{eqnarray}
\frac{d\bm{F}}{dt} = 2\pi \gamma \bm{F} \times \bm{B},
\end{eqnarray}
where $\bm{F}$ is an atomic angular momentum vector and $\bm{B} = (B_x, 0, B_z)$.
We calculated the $z$ component of $\bm{F}$ averaged over $x$ from $-75$~$\mu$m to $75$~$\mu$m with 20 segments,
$\left<F_z\right>$, by solving this equation
under the initial condition of $\bm{F} = (0,0,-|\bm{F}|)$,
as shown by the green dashed-dotted line in Fig.~{\ref{fig:results2},
where $\left<F_z\right>$ is biased and normalized to fit the height of the first Fourier component.
This calculation agrees with the experimental result much better than the Fourier spectrum.

\begin{figure}[htb]
\includegraphics[width=0.8\linewidth]{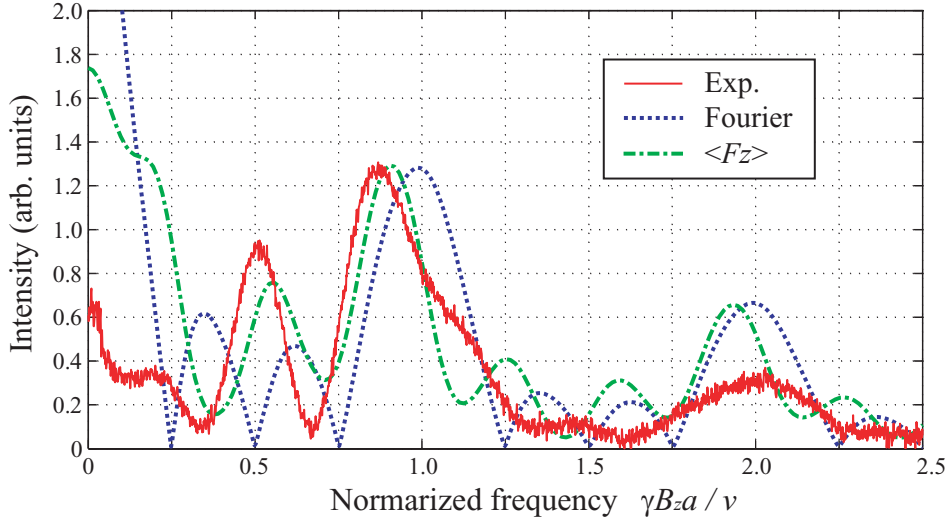}
\caption{\label{fig:results2}
As Fig.~{\ref{fig:results}}, but for the aligned magnetization direction.
}
\end{figure}

If the multi-layered transmission magnetic grating is assembled to provide a larger $N$ and shorter $a$,
its frequency band and relative linewidth are expected to be extended and narrowed, respectively.
Assembling a shorter $a$ is possible by use of thin magnetic films, as described above.
Here, we calculated $\left<F_z\right>$ for the multi-layered transmission magnetic grating with large $N$ 
to investigate the convergence of the relative linewidth.
The calculation is performed for the same dimension (see Fig.~{\ref{fig:grating}}) except for $N$.
Figures~{\ref{fig:precision}~(a) and (b) show $\left<F_z \right>$ for $N=2$ and $\mu_0 M=100$~mT (4 layers) and $N=10$ and $\mu_0 M=100$~mT (20 layers) 
gratings with alternate magnetization, respectively.
Different thin lines show the $x$ dependence for $x=-60, -30, 0, 30$ and $60$~${\mathrm{\mu m}}$
representing 5 segments in the range from $x=-75$~${\mathrm{\mu m}}$ to 75~${\mathrm{\mu m}}$.
A thick line shows the average of these lines. 
The FWHM of the thick line for $N=2$ is 0.43.
It is noted that the peak position is lower than 1 because the peak frequency depends on $N$.
For example, if $N$ cycles of the sine function are taken to be the periodic magnetic field,
the shift is described approximately by $1/(6N^2)$ calculated with the Fourier transformation.
This shift converges quickly with $1/N^2$ for larger $N$ in contrast to the relative linewidth of $1/N$.

The FWHM for $N=10$ in Fig.~{\ref{fig:precision}~(b) is expected to be five times smaller than $N=2$.
However, it is actually 0.13, which is larger than expected.
In this case, the thin lines show different shapes depending on $x$ and cause the broad linewidth.
Therefore, we calculated the Rabi frequency $\Omega = \gamma B_{xm}/2$.
$\left<F_z \right>$ is the maximum at the half period of the Rabi cycle, which becomes $\pi/\Omega = 3.2$~$\mu$s
for $B_{xm}= 0.068$~mT ($\mu_0 M=100$~mT) at $x = 0$.
On the other hand, the transit time through the grating for $N=10$ is 12~$\mu$s for a path length of $2Nd$ and $v=408$~m/s.
Therefore, the half period of the Rabi cycle is too short compared to the transit time.
To obtain a narrower linewidth, $B_{xm}$ should be lower for larger $N$ and be decreased so that $B_{xm} \propto 1/N$
to ensure that the half period of the Rabi cycle is shorter than the transit time.
The strength of $B_{xm}$ can be selected by changing the dimension of the grating, even if $a$ is fixed.
Provided that $a \lesssim s$, $B_{xm}$ at $x = 0$ is roughly described by,
\begin{eqnarray}
B_{xm} \sim \frac{2\mu_0 M\arctan(t/s)}{\pi},
\end{eqnarray}
generated by one magnetic dipole in the magnetic film, and is calculated as 0.086~mT for $\mu_0 M=100$~mT,
which is larger than the numerical calculation of 0.068~mT.
The discrepancy is caused by other magnetic dipoles present in the magnetic film, 
but is not too large to prevent estimation of the field.
From this formula, $B_{xm}$ is adjustable by $s$, $t$, and $M$ for fixed $a$.
Here, we decreased $M$ to 20~mT and calculated, as shown in Fig.~{\ref{fig:precision}}~(c).
The FWHM is 0.084, as expected.

Figures~{\ref{fig:precision}~(d) and (e) show $\left<F_z\right>$ in the aligned magnetization direction
for $N=4$ and $\mu_0 M=100$~mT (4 layers) and $N=20$ and $\mu_0 M=20$~mT (20 layers) gratings, respectively.
Lines are the same as in Fig.~{\ref{fig:precision}~(a).
The FWHMs of the thick lines in Figs.~{\ref{fig:precision}}~(d) and (e) are 0.22 and 0.043, respectively, as expected.
Although $B_b$ causes a frequency shift of $\gamma B_b^2/(2B_z)$,
this shift can be suppressed as $1/N^2$ for larger $N$,
because $B_b$ is basically proportional to $B_{xm}$, i.e., $B_{b} \propto B_{xm} \propto 1/N$.

From these measurements and calculations,
we conclude that the two types of multi-layered transmission magnetic grating can extend the frequency band by using a short $a$,
and can narrow the relative linewidth by stacking layers with large $N$ without serious frequency shifts.
Considering the ease of manufacturing the gratings, the aligned magnetization for the multi-layered transmission magnetic grating
is more efficient than the alternate magnetization.
For example, the multi-layered magnetic films can be manufactured 
by means of depositing the film with alternating layers of magnetic and non-magnetic material,
thus potentially achieving $a$ on the order of nanometers.
This realizes a frequency $v/a$ of up to hundreds of PHz for $v \sim 10^8$~m/s at a maximum.

\begin{figure}[htb]
\includegraphics[width=1.\linewidth]{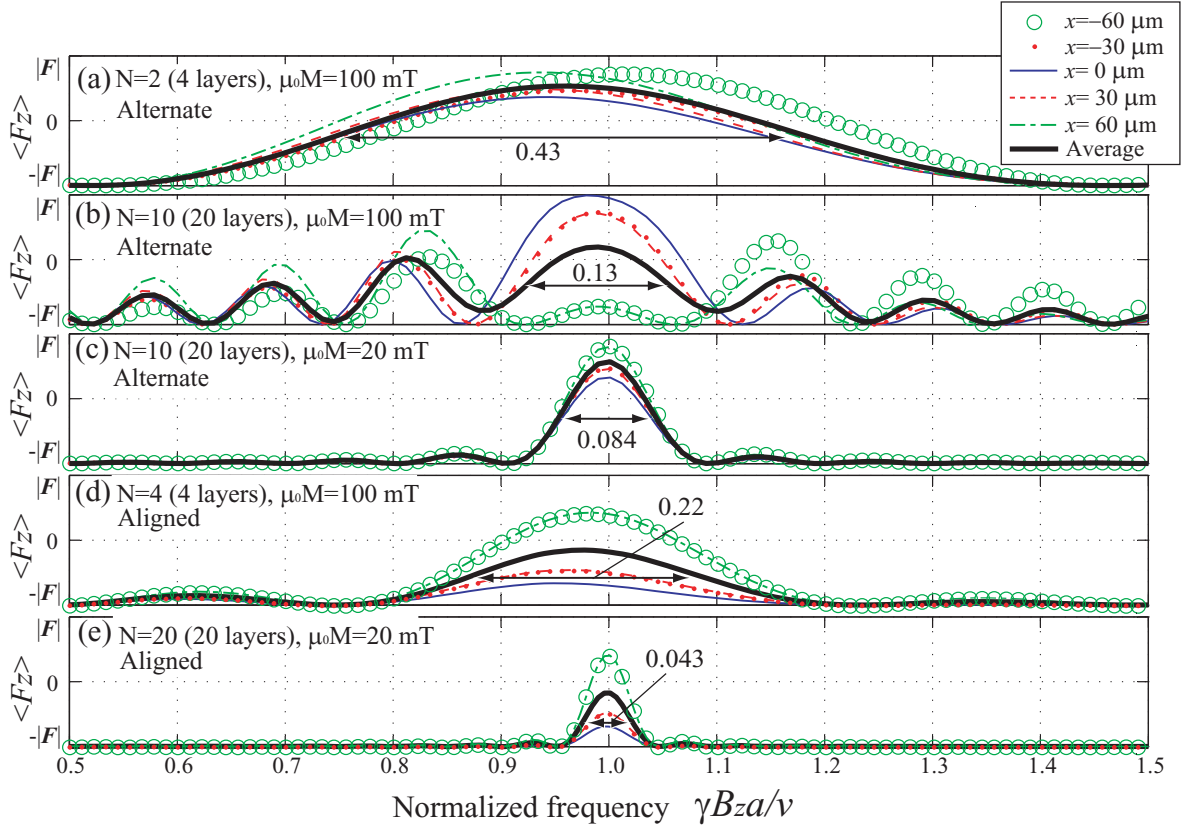}
\caption{\label{fig:precision}
(a) $\left<F_z\right>$ for $N=2$  and $\mu_0 M=100$~mT (4 layers) for the alternate direction of magnetization.
(b) $\left<F_z\right>$ for $N=10$ and $\mu_0 M=100$~mT (20 layers) for the alternate direction of magnetization.
(c) $\left<F_z\right>$ for $N=10$ and $\mu_0 M=20$~mT (20 layers) for the alternate direction of magnetization.
(d) and (e) are the same as (a) and (c) but for the aligned direction of magnetization, respectively.
}
\end{figure}

\section{\label{sec:level4}Conclusions}

We measured MIR signals of Rb atoms passing through two types of multi-layered transmission magnetic grating,
assembled in respective alternate and aligned magnetization layers,
to study the feasibility of extending the resonance frequency band of this grating and narrowing its resonance linewidth.
The MIR signal in alternate magnetization was in agreement with a Fourier spectrum.
Although the MIR signal in aligned magnetization was observed at a lower frequency than expected due to the bias field $B_b$,
the $z$ component of the atomic angular momentum vector $\left<F_z\right>$ calculated with a classical spin dynamics differential equation
reproduced the MIR signal reasonably well.

We calculated $\left<F_z\right>$ for the multi-layered transmission magnetic grating with increased $N$.
The linewidth could be narrowed by stacking more layers without serious frequency shifts.
Considering the ease of manufacturing the gratings, the aligned magnetization for the multi-layered transmission magnetic grating
is more efficient than the alternate magnetization.
We conclude that the multi-layered transmission magnetic grating
can generate periodic fields with narrower linewidths at higher frequencies by assembling a larger $N$ at a shorter $a$
for the precision spectroscopy of magnetic resonance,
and its frequency band potentially achieves frequencies of up to hundreds of PHz.

\section*{Acknowledgments}



This work was supported by JSPS KAKENHI Grant Numbers JP23244082.

\end{document}